\documentclass[preprintnumbers,amsmath,floatfix,11pt]{revtex4}
\usepackage{bm}

\begin{document}
\preprint{FERMILAB-CONF-07-48-T}
\vspace*{0.1in}

\title{Mesotron Decays and the Role of Anomalies}

\author{William A. Bardeen}

\address{Theoretical Physics Department\\
Fermilab, MS 106, P.O. Box 500\\
Batavia, IL 60510, USA\\
March 15, 2007
}

\begin{abstract}
Puzzles associated with Yukawa's mesotron theory of nuclear 
interactions led to the discovery of ``anomalies'' in quantum field
theory. I will discuss some of the remarkable consequences of these
anomalies in the physics of elementary particles.
\end{abstract}

\maketitle

In 1935 Hideki Yukawa postulated that nuclear forces were ascribed to
a new massive scalar field that coupled to neutrons to protons. To
explain the saturation of the nuclear forces, the new mesotrons were
required to have a mass of order 200 $\textrm{m}_{\textrm{e}}$, and a
coupling a few times larger than that associated with the electric
charge. The terms mesotron was used to describe a particle of
intermediate mass, much heavier than the light electron and much
lighter than the neutron and proton, the constituents of the atomic
nucleus. 

Indeed, in 1937, a new meson of intermediate mass was discovered as
the dominant part of the hard component of cosmic rays. It was natural
to associate this meson with the field that Yukawa had proposed to
explain the nuclear force. 

Shoichi Sakata played an important role in the struggle to understand
the physics of the new Yukawa mesotrons and their relation to the new
mesons seen in cosmic rays. 

Early estimates of the lifetime of the mesotron were based on Fermi's
theory of $\beta$-decay. These estimates, in the range of $10^{-8}$ to
$10^{-7}$ sec, were considerably shorter than the lifetime observed
for the cosmic ray mesons, $\sim$2 $10^{-6}$ sec. This discrepancy was
the focus of several papers by Yukawa, Sakata and
collaborators \cite{1,2}. 

Sakata also speculated on the lifetime of a neutral mesotron
(Neutretto) whose existence was suggested by the charge independence
of the nuclear force. Sakata and Tanikawa\cite{3} suggested these
neutral mesotrons should decay to photons via the following process: 
\begin{quote}
``First a neutral mesotron is absorbed by a proton which is in the
negative energy state and produces a (virtual) pair of a proton and an
antiproton. Then this pair disappears with the emission of more than
two photons'' 
\end{quote}
Their estimate the lifetime was $\sim$$10^{-16}$ sec, remarkably close
to the present measured value of the neutral pion lifetime, 
$(0.84\pm0.06)\times 10^{-16}$ sec. 

The struggle with the charged mesotron lifetime and the fact that the
cosmic ray meson did not interact strongly with nuclei led Tanikawa
and Sakata to propose in 1942 that the mesotron and the cosmic ray
meson were distinct states \cite{4}. Sakata also suggested that the
cosmic ray meson should be a fermion while the Yukawa mesotron is
required to be a boson. 

These speculations were confirmed in 1947 with the first observation
of the charged pion in high altitude cosmic rays \cite{5} and later by
pions produced artificially by accelerators \cite{6} in 1948. The
neutral pion was also discovered in its two-photon decay mode at
accelerators \cite{7} in 1950. 

However, theoretical estimates for the lifetime of the neutral pion
would lead to another puzzle. New field theory methods were developed
to perform the calculation of meson lifetime along the lines
originally suggested by Sakata that the neutral meson decays to
photons via virtual proton loop. These new methods gave a finite
result for a pion with pseudoscalar coupling and a lifetime estimate,
$\sim$0.9 $10^{-16}$ sec. However, the axial-vector coupling remained
divergent \cite{8}. The calculations were plagued with divergences and
questions of electromagnetic gauge invariance. 

A regularization method developed by Pauli and Villars\cite{9} using
heavy fermion regulators would permit the gauge invariant calculation
of both pseudoscalar and axial-vector matrix elements. This
calculation by Steinberger\cite{10} concluded that the pseudoscalar
coupling and the axial-vector coupling of the pion to the proton gave
finite but different results for the decay amplitude. The field theory
equations of motion should have implied that the amplitudes were the
same \cite{11} leading to the puzzle concerning the actual prediction
for the neutral pion lifetime. 

The axial-vector current gained significance with the observation that
the strongly interacting particles possess an approximate, hidden
chiral symmetry in addition to isospin. Although the nucleon mass
appears to break chirality, axial-vector current remains conserved due
to the existence of massless pions, PCAC. The pion is a
Nambu-Goldstone boson, 
\begin{equation*}
\left<p^{\prime}\left|J_{5\lambda}^{}\right|p\right>
=\overline{u}(p^{\prime})
\left\{g_{A}^{}\gamma_{5}^{}\gamma_{\lambda}^{}\right\}u(p)
+f\left(q_{\lambda}^{}/q^{2}\right)\overline{u}(p^{\prime})
\left\{g_{P}^{}\gamma_{5}^{}\right\}u(p),\quad
q=p^{\prime}-p
\end{equation*}
and the axial-vector current remains conserved if
$g_{A}^{}2M=fg_{P}^{}$, the Goldberger-Treiman relation\cite{12}. The
pion pole compensates for lack of conservation of the axial-vector
form factor contribution. The enhanced role of axial-vector current is
a reflection of the chiral symmetry of the equations of motion. 

An explicit realization of PCAC is provided by the Gell-Mann-L\'{e}vy
sigma model \cite{13}, a renormalizable model of pions, nucleons and a
scalar meson. Chiral symmetry is dynamically broken by a scalar
condensate or vev. A term linear in the scalar field can be added to
the model which explicitly breaks the chiral symmetry and generates a
small pion mass. In this case, the divergence of the axial-vector
current is exactly proportional to the pion field. This model is
described by the Lagrangian, 
\begin{equation*}
L_{I}^{}=-\left(m/f\right)\overline{N}
\left(\sigma+i\vec{\tau}\cdot\vec{\pi}\gamma_{5}^{}\right)N
-\lambda\left(f^{2}-\sigma^{2}-\vec{\pi}^{2}\right)^{2}+f\mu^{2}\sigma.
\end{equation*}

In 1969, Bell and Jackiw revisited the calculation of the two-photon
matrix element of the axial-vector current in light of PCAC and its
explicit realization in the Gell-Mann-L\'{e}vy sigma model. Using the
Steinberger\cite{10} calculation of the nucleon loops, they found a
finite result but that PCAC is explicitly violated for photon matrix
elements, the Bell-Jackiw anomaly \cite{14}. They suggest that the
fault lies with the use of the Pauli-Villars regulators for the loop
calculations and propose an alternative, chiral invariant regulator
where the coupling of pions to the heavy regulator fermions grows with
the regulator mass. PCAC is restored in the regulated theory but the
decay of the neutral pion to photons is highly suppressed. The
regulator fields interact according to 
\begin{equation*}
L_{R}^{}=-\left(M_{R}^{}/f\right)\overline{\psi}_{R}^{}
\left(\sigma+i\vec{\tau}\cdot\vec{\pi}\gamma_{5}^{}\right)\psi_{R}^{}.
\end{equation*}

However, the regulator loops do not decouple for large regulator mass
and additional nonrenormalizable interactions are generated including
a counterterm to the pion-photon-photon vertex. These new interactions
will become known as Wess-Zumino terms in nonlinear realizations of
dynamical chiral symmetry breaking. 

At this same time, Adler studied the properties of the axial-vector
current in spinor electrodynamics, a renormalizable model with a
chiral symmetry that is only broken by the fermion mass
term. Electromagnetic gauge invariance and power counting completely
specify matrix elements of the axial vector current. The axial-vector
current has an unambiguous anomalous divergence associated with the
two photon matrix elements. The field equations for the divergence of
the axial current contain a specific additional contribution, the
Adler anomaly \cite{15}. 
\begin{equation*}
\partial^{\mu}\left\{\overline{\psi}\gamma_{\mu}^{}\gamma_{5}^{}\psi\right\}
=2m\left\{\overline{\psi}i\gamma_{5}^{}\psi\right\}
+\frac{\alpha}{4\pi}Q^{2}F^{\mu\nu}\cdot{}^{*}\!F_{\mu\nu}^{}
\end{equation*}
The anomaly coefficient is determined by the fundamental charge of the
fermion. There is an exact low energy theorem for matrix element of
the na$\ddot{\i}$ve divergence of the axial-vector current. This low
energy theorem is not modified to all orders in perturbation theory,
the Adler-Bardeen nonrenormalization theorem \cite{16}. This
nonrenormalization theorem identifies the anomaly as a fundamental
aspect of quantum field theory and not simply an artifact of a
particular perturbative calculation. 

The anomalous divergence of the axial-vector isospin current
determines the low energy theorem for neutral pion decay to two
photons. Since this anomaly can be computed exactly in a specific
model, it can be used to identify consistent models of the strong
dynamics. The observed lifetime of the pion disagrees with the
uncolored quark model \cite{15} and provides evidence for the color
triplet quark model \cite{17}, 
\begin{equation*}
\begin{array}{lcl}
\textrm{Nucleon Model (Steinberger):}&1^{2}-0^{2}&=1\\
\textrm{Triplet Quark Model:}&(2/3)^{2}-(-1/3)^{2}&=1/3\\
\textrm{Color Triplet Quark Model:}&3(2/3)^{2}-3(-1/3)^{2}&=1
\end{array}
\end{equation*}

Anomalies are not restricted to the abelian case considered by Adler,
Bell and Jackiw. Anomalies can be generalized through the study of
nonabelian currents in field theory. Nonabelian currents can be
derived from a general fermion Lagrangian with arbitrary couplings to
vector, axial vector, scalar and pseudoscalar fields, 
\begin{equation*}
L=\overline{\psi}
\left\{\gamma^{\mu}V_{\mu}^{}+\gamma^{\mu}\gamma_{5}^{}A_{\mu}^{}
-\Sigma-i\gamma_{5}^{}\Pi\right\}\psi
=\overline{\psi}\left\{\Gamma\right\}\psi
\end{equation*}
Anomalous terms are generated by considering local gauge
transformations involving all fields in the effective action which is
formally gauge covariant. A precise study shows that all of the
anomalous terms can be made to cancel except for those involving
certain external vector and axial vector fields with the
result \cite{18}, 
\begin{eqnarray*}
D\left(\Lambda_{+}^{},\Gamma\right)&=&R\left(\Gamma
i\Lambda_{+}^{}-i\Lambda_{-}^{}\Gamma
-\gamma\cdot\partial\Lambda_{+}^{},\Gamma\right)\\
&=&\frac{1}{6}\frac{\pi^{2}}{(2\pi)^{4}}\,i\!\int\!
dz\,\epsilon_{\mu\nu\sigma\tau}^{}\textrm{tr}\gamma_{5}^{}
\left\{2i\Lambda_{+}^{}\partial^{\mu}V_{+}^{\nu}\partial^{\sigma}V_{+}^{\tau}
-\partial^{\mu}V_{+}^{\nu}V_{+}^{\sigma}V_{+}^{\tau}\right\}
\end{eqnarray*}
where $V_{+}^{v}=V^{v}+A^{v}\gamma_{5}^{}$,
$\Lambda_{+}^{}=\Lambda+\Lambda_{5}^{}\gamma_{5}^{}$. 

It is remarkable that the only anomalous divergences that survive are
those associated with the gauge fields themselves. Indeed, the fermion
loops can be defined such that the anomalous divergence of left-handed
currents involves only left-handed gauge fields. The fermion loops can
also be redefined so that all vector currents are conserved and only
axial-vector currents have anomalies \cite{18}. Anomalies reflect
classical symmetries that clash at the quantum level. 

An intuitive perspective of fermion loop anomalies was provided by
Fujikawa\cite{19} who showed that anomalies could be viewed as
resulting from the variation of the fermionic measure in a path
integral formulation of quantum field theory. This view became of
great importance in implementing fermions in lattice field theory and
the study of the precise realization of chiral symmetry on the
lattice \cite{20}. 

The nonrenormalization theorem states that radiative corrections do
not modify the fermion loop anomaly and that the only sources of
anomalies are the small fermion loops. The Adler-Bardeen
nonrenormalization theorem can be generalized to arbitrary
renormalizable quantum field theories by using explicit regularization
methods \cite{21} or renormalization group arguments \cite{22}. 

As discussed above anomalies can be viewed as arising from the local
gauge variation of the non-local effective action describing the
fermion loop dynamics. Since two gauge transformations are again a
gauge transformation, the explicit functional form of anomalies is
constrained by the consistency, or integrability, conditions of Wess
and Zumino\cite{23}. The full nonabelian anomaly of Bardeen\cite{18}
was shown to satisfy these consistency conditions. 

Wess and Zumino\cite{23} also showed that the consistency conditions
could be integrated using an effective action involving only nonlinear
realizations of Nambu-Goldstone boson fields. This effective action
defines a set of Wess-Zumino terms that reproduce anomalous amplitudes
for the nonabelian currents. In this case anomalies are related to the
propagation of massless bosons and not fermion loops. Indeed, the
chiral invariant regulator of Bell and Jackiw\cite{14} can be viewed,
in part, as generating additional Wess-Zumino terms which are then
used cancel the fermion loop anomalies in their chiral invariant
version of the Gell-Mann-L\'{e}vy sigma model. 

Anomalies have had a remarkable impact on developments in particle
physics, string theory and condensed matter physics. Classical
applications to particle physics include the role of anomaly
cancellation on building consistent gauge theories, the impact of
anomalies on global symmetries and the constraints anomalies impose on
nonperturbative dynamics in field theory. 

Since anomalies reflect the quantum breaking of classical gauge
symmetries, all anomalies associated with the dynamical gauge currents
must cancel for the gauge dynamics to be consistent at the quantum
level. Vector-like gauge theories such as electrodynamics and quantum
chromodynamics are automatically free of dynamical anomalies since all
of the potential anomalies can be removed through the appropriate
choice of counterterms \cite{18}. However, anomaly cancellation
constrains the matter content of chiral gauge theories. Electroweak
gauge theories do have chiral couplings and will generally be expected
to have potential anomalies. However, all electroweak anomalies  are
seen to cancel in the Standard Model of particle physics\cite{24}. 
\begin{equation*}
\begin{array}{cccc}
\textrm{Standard Model}&\textrm{Leptons}&\textrm{Quarks}&\textrm{Sum}\\
SU(2)^{2}\bigotimes
U(1)&-1/2&3\cdot(1/6)&0\\
U(1)^{3}&1-1/4&1/36-8/9+1/9&0
\end{array}
\end{equation*}
Lepton anomalies do not vanish but are canceled by the quark anomalies
in each generation of fermions. A new puzzle arises: Is this
remarkable cancellation an accident or a reflection of a deeper
connection between quarks and leptons such as grand unification,
compositeness, or some other property of the fundamental dynamics?
Anomalies continue to play a central role in building models to
understand the potential for physics beyond that described by the
present Standard Model. 

Anomalies can also reflect a clash between global symmetries and the
dynamical gauge symmetries. The electromagnetic gauge fields were seen
to add an anomalous term to the divergence of the singlet axial vector
current in the case of the Adler-Bell-Jackiw anomaly. Similarly, the
color gauge fields of quantum chromodynamics generate an anomalous
contribution to the U(1) axial vector current in QCD. This anomaly
provides the potential for a solution to the U(1) problem of QCD: why
the singlet eta' meson is not approximately degenerate with the
neutral pion. 

In a remarkable paper, 't Hooft\cite{25} made a precise calculation of
the effects of nonperturbative gauge fields in the form of
pseudoparticles, or instantons, on anomalous symmetries. He showed
that these nonperturbative effects do generate explicit breaking of
the anomalous symmetries, such as baryon number violating processes in
the Standard model. 

Similar effects provide a possible solution of the U(1) problem in
QCD\cite{26} and a large mass for the eta' meson. However, this
solution also generates a new puzzle in the form of a new CP violating
parameter of the strong dynamics, the theta angle, which could be of
order 1 but is highly constrained by experiment to be tiny,
$\theta<10^{-9}$. This constraint on the size of the theta angle
represents an outstanding fine-tuning problem of QCD and the Standard
model of particle physics. 

A possible solution to this fine tuning problem involves the dynamical
breaking of a new Peccei-Quinn symmetry \cite{27} that allows the theta
angle to relax to zero. This dynamical breaking also introduces a new
Nambu-Goldstone boson, the axion\cite{28}. Despite the role of
nonperturbative QCD dynamics, detailed predictions \cite{29} for the
phenomenology of axions could be made using anomalous current algebra
with the result that electroweak scale axions could be ruled out. The
strong CP problem is still outstanding for QCD and the Standard Model,
and the search for an ``invisible'' axion continues. 

As mentioned above, electroweak instanton effects associated with the
electroweak gauge field can directly induce baryon number violation
processes in the Standard Model. Although highly suppressed in the our
present vacuum state, these effects could be important at the high
temperatures present near the electroweak phase transition in the
early universe. Baryon number asymmetries could also be generated via
leptogenesis. In this case lepton number violating processes in the
early universe can be converted to baryon asymmetries by the anomalous
B+L violating processes of the Standard Model. 

Anomalies can also occur through the clash between purely global
symmetries. While anomalies are a fundamental aspect of the
short-distance dynamics, the infrared realization of the fundamental
global anomalies may depend on the details of the low energy
dynamics. Consistency conditions can be derived \cite{30} to relate the
infrared and ultraviolet behavior of anomalies which implies
constraints on boundstate structure in composite models, anomalous
couplings, etc. In QCD, the quark picture applies in ultraviolet but
pions carry the QCD dynamics in infrared including the effects of the
global anomalies. 

The low energy theorems of current algebra can be encoded in effective
field theories describing the relevant physics at low energies. Global
anomalies will be reflected by specific anomalous terms in the
effective Lagrangians of the low energy dynamics. Wess and Zumino
showed that fermion loop anomalies could be reproduced using specific
nonlinear realizations of the global current algebra and encoded in
Wess-Zumino terms \cite{31}. In the case of chiral symmetry, the full
set of Wess-Zumino terms predicts additional anomalous multipion interactions
including the anomalous coupling of pions to external gauge fields and
currents. An elegant formulation of these anomalous interactions,
including the coupling to gauge fields, was made by Witten\cite{32}
and extended by others \cite{33}. 

Anomalies can also be discussed within the context of the local
symmetries associated with gravity and external gravitational
fields. Kimura\cite{34} showed that U(1) axial-vector anomaly contains
an additional contribution in the presence of background gravitational
fields.
\begin{equation*}
\partial^{\mu}J_{5\mu}^{}=\frac{1}{768\pi^{2}}\epsilon_{\mu\nu\sigma\tau}^{}R^{\mu\nu\alpha\beta}
R_{\alpha\beta}^{\sigma\tau}
\end{equation*}
This contribution is related to a topological index for the
gravitational field, similar to the gauge anomaly where 
the anomalous divergence is related to a topological index 
for the nonabelian gauge field. The consistency
of the U(1) gauge interactions in the Standard Model implies that the
sum of all contributions of the individual fermions to the
gravitational anomaly must cancel. Another remarkable feature of the
Standard Model is that this cancellation, in fact, automatically
occurs within each generation. 
\begin{equation*}
\begin{array}{cccc}
\textrm{Standard Model}&\textrm{Leptons}&\textrm{Quarks}&\textrm{Sum}\\
R^{2}\bigotimes
U(1)&2(-1/2)+1&3(1/3)+3(-2/3)+6(1/6)&0
\end{array}
\end{equation*}
In this case, the lepton and quark gravitational anomalies cancel
independently. Can this fact provide any additional hints for the
specific origin of the fermion representations used by nature? 

Pure gravitational anomalies also exist in 2, 6 and 10
dimensions \cite{35}. These anomalies could imply the existence of
possible obstructions to the formulation consistent gravitational
couplings in these dimensions. Gravity has many local symmetries
including general coordinate invariance and the local Lorentz
invariance of the tangent space required to define fermions. The
consistency of gravitational anomalies and the connection between
anomalies associated with general coordinate symmetries and the
gauging of the local Lorentz symmetries can be derived \cite{36}. 

Anomalies have also played an important role in the development of
string theory. String theories must avoid both gauge and gravitational
anomalies. Anomalies can be seen by studying the massless sector of
the theory as the anomalous low energy theorems require a
nonanalyticity of the relevant amplitudes. For example, the massless
sector of open string states can contribute to anomalous amplitudes at
one loop where the relevant terms are described by a cylinder with a
large radius but small height. The cylinder function has another
infrared limit where a closed string tube connects two disks where the
tube has a small radius and large length. In this limit, the anomalous
behavior arises at tree level from massless poles in the gravity
sector of the theory. Green and Schwarz\cite{37} made the crucial observation
that loop anomalies found in the open string sector could cancel
against the tree level anomalous couplings of gravitational sector
with both contributions arising from the same cylinder
contribution. With this observation modern string theory was born. It
is amusing to observe that this mechanism for the cancellation of
string anomalies is similar to the anomaly cancellation seen by Bell
and Jackiw in their chiral invariant regularization of the
Gell-Mann-L\'{e}vy sigma model or in the Standard Model between pions
and lepton loops. 

\section*{Conclusions}

Sakata's detailed study of mesotron decays led him to speculate about
the decay of the neutral mesotron to photons. New technologies in
quantum field theory allowed the precise calculation of the neutral
pion decay amplitudes following the ideas suggested by Sakata but left
puzzles concerning the discrepancies between pseudoscalar and
axial-vector couplings of the pion. 

The suggestion of a hidden chiral symmetry of the strong interactions
focused attention on the role of the axial-vector current and a new
realization of symmetries in the form of PCAC with the role of the
pion as a Nambu-Goldstone boson. The extension of PCAC to amplitudes
involving photons led to the discovery of anomalies and a more
fundamental understanding of quantum field theories and their
symmetries. 

The discovery and analysis of the complete nonabelian anomaly coupled
with nonrenormalization theorems showed that anomalies reflect very
fundamental aspects of quantum field theory. Anomalies have a
remarkably broad impact on theoretical particle physics, from anomaly
cancellation as an intrinsic part of model building to determining
subtle aspects of realizations of global symmetries and constraints on
the nonperturbative dynamics. Anomalies also play an important role in
modern string theory and can be expected to be an essential part of
our ultimate description of nature. 

\section*{Acknowledgements}

Talk presented at the 2006 International Symposium of the Jubilee of 
the Sakata Model (pn$\Lambda$50), held November 25-26, 2006 at Nagoya, 
Japan.  I would like to thank Prof. Koichi Yamawaki and other members 
of the Organizing Committee for inviting me to participate in the 
Sakata Jubilee Symposium. 

Fermilab is operated by Fermi Research Alliance,
LLC under Contract No. DE-AC02-07CH11359 with the United States
Department of Energy.


\begin{thebibliography}{99}
  
\bibitem{1}
S. Sakata and H. Yukawa, Proc. Phys. -Math. Soc. \textbf{21} (1939), 138; ibid, Nature \textbf{143} (1939), 761.

\bibitem{2}
S. Sakata, Phys. Rev. \textbf{58} (1940), 576; ibid, Proc. Phys. -Math. Soc. \textbf{23} (1941), 283; ibid, Proc. Phys. -Math. Soc. \textbf{23} (1941), 291; ibid, Proc. Phys. -Math. Soc. \textbf{24} (1942), 843.

\bibitem{3}
S. Sakata and Y. Tanikawa, Phys. Rev. \textbf{57} (1940), 548.

\bibitem{4}
H. Yukawa, Nobel Lecture, 1949; S. Sakata and K. Inouye, Prog. Theor. Phys. \textbf{1} (1946), 143; Y. Tanikawa, Prog. Theor. Phys. \textbf{2} (1947), 220; R. Marshak and H. Bethe, Phys. Rev. \textbf{72} (1947), 506.

\bibitem{5}
C. Lattes, H. Muirhead, G. Occhialini and C. Powell, Nature \textbf{159} (1947), 694; C. Lattes, G. Occhialini and C. Powell, Nature \textbf{160} (1947), 452, 486.

\bibitem{6}
E. Gardner and C. Lattes, Science \textbf{107} (1948), 270; W. Barkas, E. Gardner and C. Lattes, Phys. Rev. \textbf{74} (1948), 1558; J. Richardson, Phys. Rev. \textbf{74} (1948), 1720.

\bibitem{7}
J. Steinberger, W. Panofsky and J. Steller, Phys. Rev. \textbf{78} (1950), 802; H. York, B. Moyer and R. Bjorklund, Phys. Rev. \textbf{76} (1949), 187.

\bibitem{8}
H. Fukuda and Y. Miyamoto, Prog. Theor. Phys. \textbf{4} (1949), 347.

\bibitem{9}
W. Pauli and F. Villars, Rev. Mod. Phys \textbf{21} (1949), 434.

\bibitem{10}
J. Steinberger, Phys. Rev. \textbf{76} (1949), 1180.

\bibitem{11}
J. Schwinger, Phys. Rev. \textbf{82} (1951), 664.

\bibitem{12}
M. Goldberger and S. Treiman, Phys. Rev. \textbf{110} (1958), 1178.

\bibitem{13}
M. Gell-Mann and M. L\'{e}vy, Nuovo. Cim. \textbf{16} (1960), 705.

\bibitem{14}
J. Bell and R. Jackiw, Nuovo. Cim. \textbf{60} (1969), 47.

\bibitem{15}
S. Adler, Phys. Rev. \textbf{177} (1969), 2426.

\bibitem{16}
S. Adler and W. Bardeen, Phys. Rev. \textbf{183} (1969), 1517.

\bibitem{17}
W. Bardeen, H. Fritzsch and M. Gell-Mann (1971-72).

\bibitem{18}
W. Bardeen, Phys. Rev. \textbf{184} (1969), 1848.

\bibitem{19}
K. Fujikawa, Phys. Rev. \textbf{D21} (1980), 2848.

\bibitem{20}
P. Ginsparg and K. Wilson, Phys. Rev. \textbf{D25} (1982), 2649.

\bibitem{21}
W. Bardeen, Proc. Marseilles Conf. on Renormalization (Marseilles, 1972) and Proc. of  the XVI Int. Conf. on HEP (Fermilab, 1972).

\bibitem{22}
A. Zee, Phys. Rev. Lett. \textbf{29} (1972), 1198.

\bibitem{23}
J. Wess and B. Zumino, Phys. Lett. \textbf{B37} (1971), 95.

\bibitem{24}
C. Bouchiat, J. Iliopoulos and P. Meyer, Phys. Lett. \textbf{B38} (1972), 519; D. Gross and R. Jackiw, Phys. Rev. \textbf{D6} (1972), 477.

\bibitem{25}
G. 't Hooft, Phys. Rev. Lett. \textbf{37} (1976), 8; Phys. Rev. \textbf{D14} (1976), 3432.

\bibitem{26}
E. Witten, Nucl. Phys. \textbf{B156} (1976), 269.

\bibitem{27}
R. Peccei and H. Quinn, Phys. Rev. Lett. \textbf{38} (1977), 1440.

\bibitem{28}
F. Wilczek, S. Weinberg, Ben Lee Memorial Conf. (1977); S. Weinberg, Phys. Rev. Lett. \textbf{40} (1978), 223; F. Wilczek, Phys. Rev. Lett. \textbf{40} (1979), 279.

\bibitem{29}
W. Bardeen and H. Tye, Phys. Lett. \textbf{B74} (1978), 229.

\bibitem{30}
G. 't Hooft, Cargese Lectures (1979); Y. Frishman, A. Schwimmer, T. Banks and S. Yankielowicz, Nucl. Phys. \textbf{B177} (1981), 157; S. Coleman and B. Grossman, Nucl. Phys. \textbf{B203} (1982), 205.

\bibitem{31}
J. Wess and B. Zumino, Phys. Lett. \textbf{B37} (1971), 95.

\bibitem{32}
E. Witten, Nucl. Phys. \textbf{B233} (1983), 422.

\bibitem{33}
O. Kaymahcalan, S. Rajeev and J. Schechter, Phys. Rev. \textbf{D30} (1984), 594; A. Manohar and G. Moore, Nucl. Phys. \textbf{B243} (1984), 55; H. Kawai and H. Tye, Phys. Lett. \textbf{B140} (1984), 403.

\bibitem{34}
T. Kimura, Prog. Theor. Phys. \textbf{42} (1969), 1191; R. Delbourgo and A. Salam, Phys. Lett. \textbf{B40} (1972), 381.

\bibitem{35}
L. Alvarez-Gaum\'{e} and E. Witten, Nucl. Phys. \textbf{B234} (1984), 269; O. Alvarez, I.M. Singer and B. Zumino, Comm. Math. (1984).

\bibitem{36}
W. Bardeen and B. Zumino, Nucl. Phys. \textbf{B244} (1984), 421.

\bibitem{37}
J. Schwarz and M. Green, Phys. Lett. \textbf{149} (1984), 117.

\end{thebibliography}
\end{document}